\documentstyle[11pt,newpasp,twoside,epsf]{article}

\newcommand{\etal}{{et al.~}}

\begin{document}

\title{The Central Kiloparsec-Scale Structure of Galaxies}

\author{Roeland P.~van der Marel}
\affil{Space Telescope Science Institute, 3700 San Martin Drive, 
Baltimore, MD 21218, USA}

\begin{abstract}
This review summarizes some aspects of the central kiloparsec scale
structure of galaxies, and in particular spiral galaxies, elliptical
galaxies and merger remnants. The focus is on results from optical and
near-IR imaging and spectroscopy, with emphasis on recent work with
the Hubble Space Telescope.
\end{abstract}

\section{Introduction}

This paper provides an overview of some aspects of the structure of
galaxies in the central kiloparsec, as relevant in the context of the
symposium topic: `Galaxies and Their Constituents at the Highest
Angular Resolutions'.

For the purpose of this introduction, let us consider high angular
resolution to mean $\sim 0.1$ arcsec or better.  This angular scale
corresponds to a linear scale of 1 pc at a distance of 2 Mpc. While
many important topics can be studied on the parsec scale in galaxy
centers (see other reviews in this volume), such studies are limited
to only a small fraction of the observational tools and astronomical
targets available to astronomers. In particular, one must either: (a)
look at very nearby galaxies (e.g., in the Local Group); (b) use the
limited set of observational techniques that provide resolutions $\ll
0.1$ arcsec (e.g., very long baseline radio interferometry); or (c)
study the global properties of regions that are not spatially resolved
(e.g., AGN broad-line regions of accretion disks).

By contrast, structures in the central kiloparsec of galaxies are
easily accessible except for very distant targets. For $0.1$ arcsec to
correspond to 1 kpc one must go to a distances of 2 Gpc, corresponding
to a redshift $z \approx 0.5$. Hence, a complete review of the central
kiloparsec scale structure of galaxies would necessarily have to
encompass a discussion of the full range of observational techniques
(imaging, spectroscopy, polarimetry, interferometry, etc.) and
wavelength ranges (radio, sub-mm/mm, far/mid-IR, near-IR/optical, UV,
X-ray, $\gamma$-ray) available, for the different structural and
physical components (stars, gas, dust, black holes, dark matter, etc.)
that are present in the large variety of interesting galaxy types in
the Universe (various Hubble types, galaxy mergers and interactions,
starburst galaxies, ultra-luminous IR galaxies, dwarf galaxies,
low-surface brightness galaxies, LINERS, Seyfert galaxies, radio
galaxies, quasars, BL Lac objects, blazars, etc.). The main challenge
is then to put the vast amounts of observational information together
into one coherent picture of galaxy structure and evolution. However,
such a detailed discussion is far beyond what is feasible in the
context of a brief review. So instead I focus here on a small
sub-topic, namely results from imaging and spectroscopy in the optical
and near-IR wavelength regimes, in particular from recent work with
the Hubble Space Telescope (HST). I describe some new insights into
the central kiloparsec scale structure of spiral galaxies, elliptical
galaxies and merger remnants.

\section{Spiral Galaxies}

\subsection{Bulges}

The central regions of spiral galaxies have traditionally been well
studied at kiloparsec scales using ground-based observations. Results
from such studies have shown that spiral galaxies have bulges, and
these bulges diminish progressively in prominence from early-type
spirals to late-type spirals. This progression forms one of the bases
of the Hubble sequence classification. The majority of early-type
spirals (S0a--Sab) have bulges with brightness profiles that follow an
$R^{1/4}$ law. Bulges of this kind have traditionally been viewed as
`small ellipticals' (e.g., Bender, Burstein \& Faber 1992). By
contrast, most intermediate (Sb--Sc) type spirals have bulges that
follow an exponential profile, and are of lower surface brightness and
density than their $R^{1/4}$ counterparts. In late type spirals (Scd
and later type) virtually all bulges are of this exponential type
(e.g. Andredakis \etal 1995; Courteau, de Jong \& Broeils 1996). The
physical processes that govern the formation of the different types of
bulges remain a topic of debate; see, e.g., the review by Wyse,
Gilmore \& Franx (1997) and the proceedings of a recent STScI
conference (Carollo, Ferguson \& Wyse 1999).


\begin{figure}[t!]
\centerline{ 
\noindent\hfill
\epsfxsize=0.3\hsize
\epsfbox{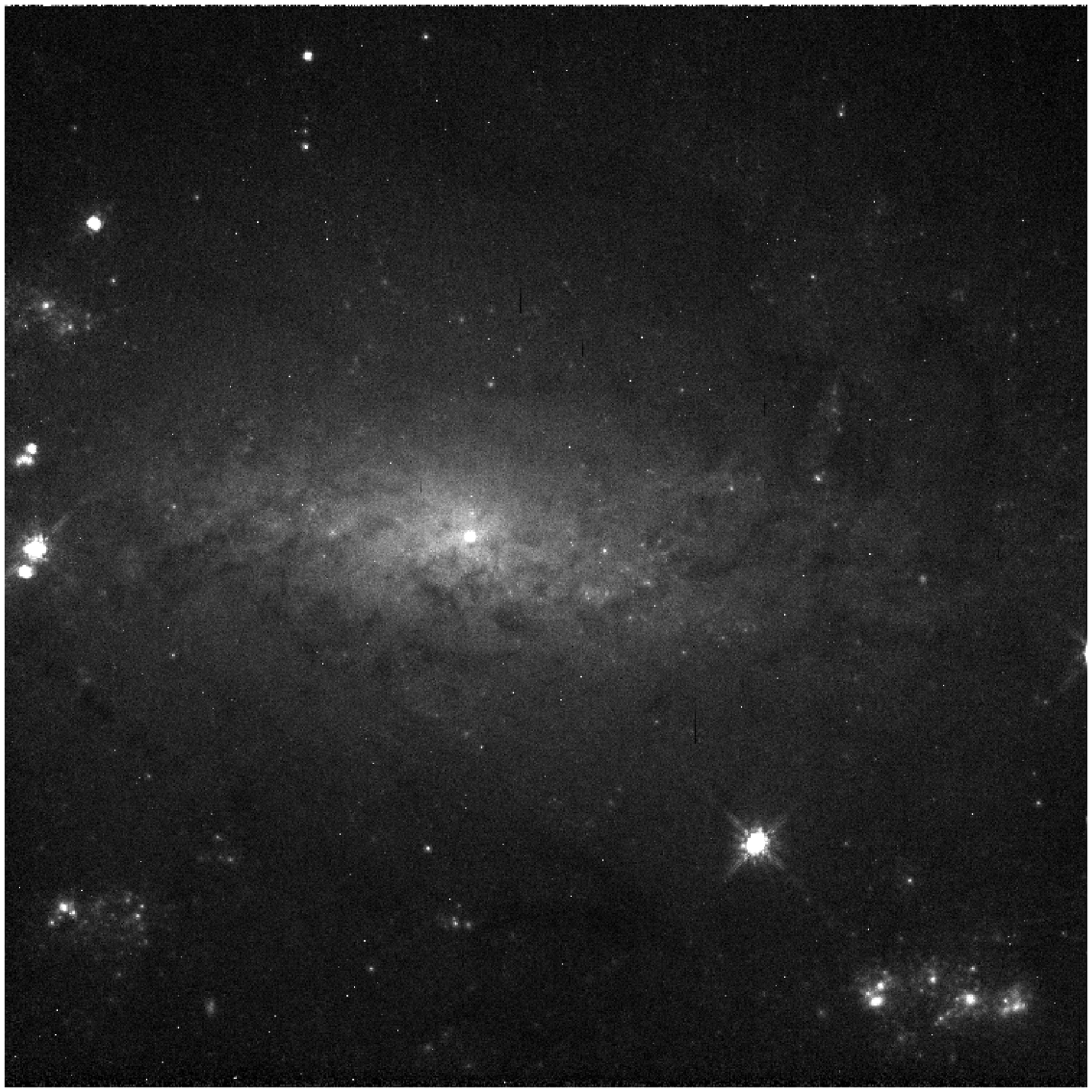}
\quad
\epsfxsize=0.3\hsize
\epsfbox{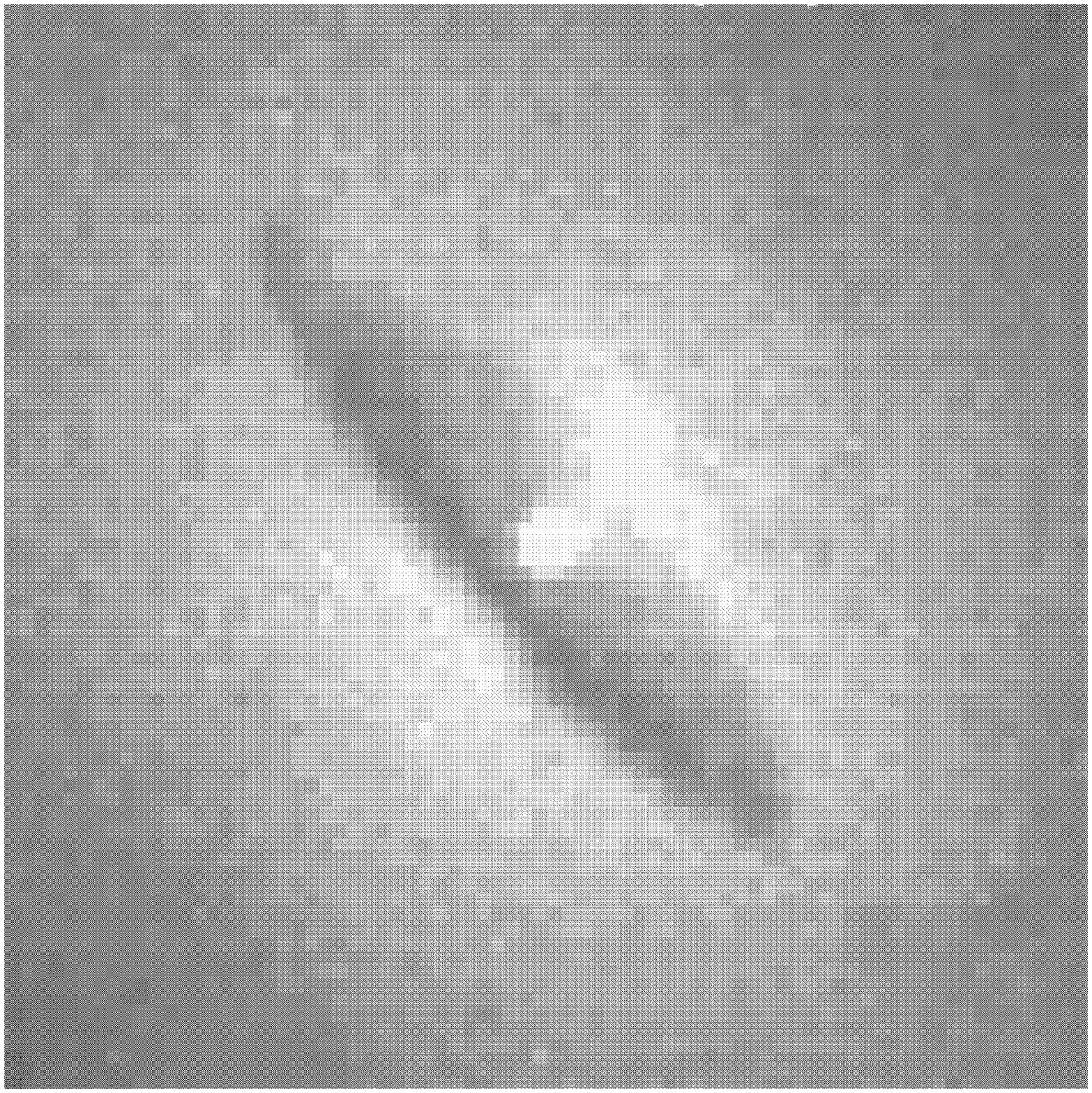}
\quad
\epsfxsize=0.3\hsize
\epsfbox{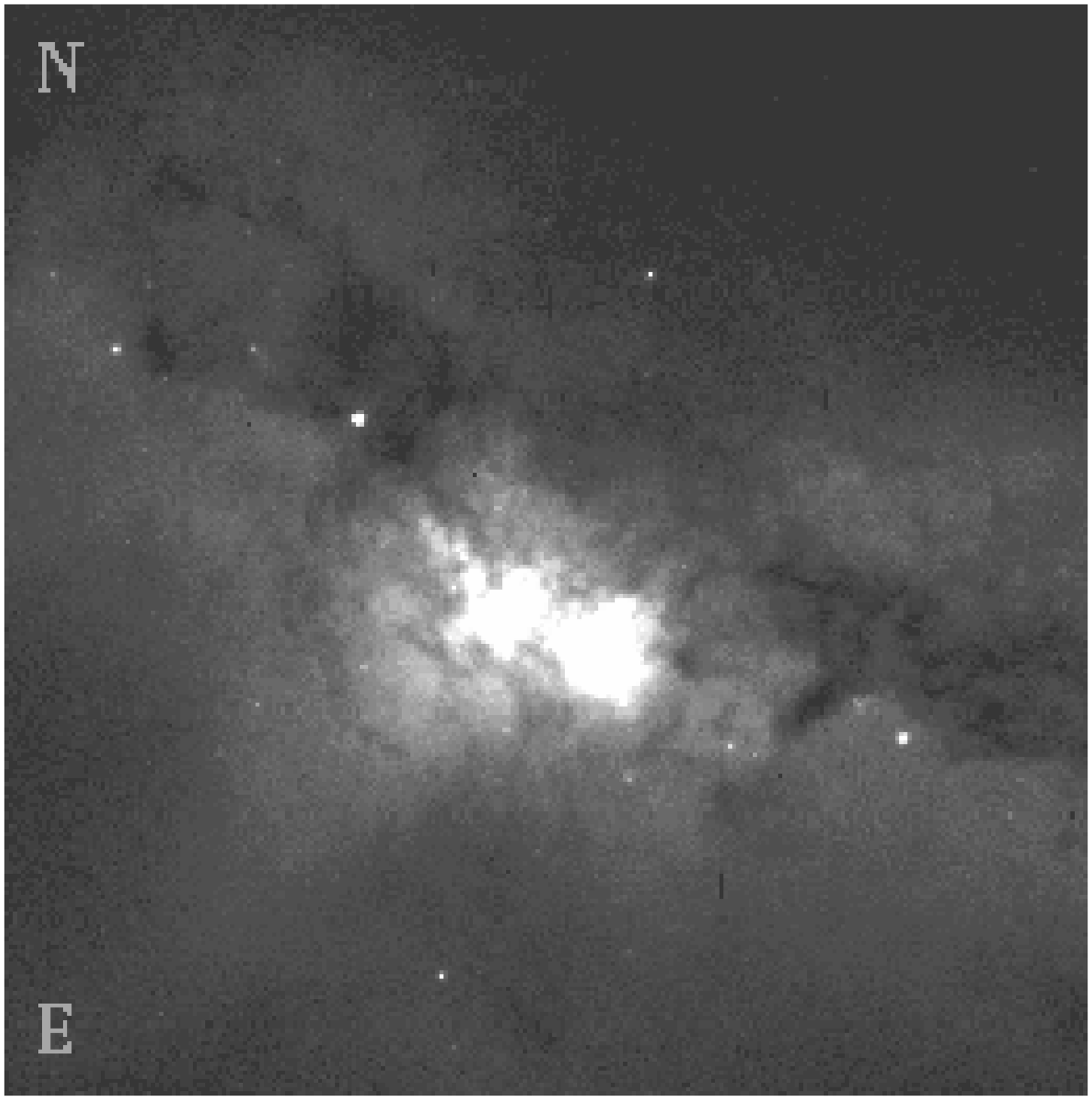}
\hfill} 
\caption{(a) HST/WFPC2 $I$-band image of NGC 6509, a late-type spiral
galaxy with a nuclear star cluster (see Section~2.2; B\"oker \etal
2001b). (b) HST/WFPC2 $V$-band image of NGC 315, an elliptical galaxy
with a nuclear dust disk (see Section~3.2; Verdoes Kleijn \etal
1999). (c) HST/WFPC2 $I$-band image of NGC 6240, a ULIRG and merger
remnant with an obscured AGN (see Section~4.1; van der Marel \etal
2001).}
\end{figure}


\subsection{Nuclear Star Clusters}

Recent HST observations at both optical and near-IR wavelengths (e.g.,
Carollo, Stiavelli \& Mack 1998; B\"oker \etal 1999a) have revealed
that a compact, photometrically distinct star cluster is often present
in the dynamical center of spiral galaxies of all Hubble types (an
example is shown in Fig.~1a).

To understand the origin and nature of these star clusters it is
desirable to know their ages. One possible approach is to attempt to
measure the stellar velocity dispersion of the cluster from high
spectral resolution ground-based absorption line spectra. We have
succeeded in doing this from the ground (with IRTF) for one very
nearby galaxy, IC 342 ($D \approx 2$ Mpc). This yields the mass of the
cluster ($\sim 6 \times 10^6 M_{\odot}$), and hence the $M/L$ from
comparison with the observed luminosity. Stellar population synthesis
models for the observed $M/L$ then yield the cluster age. For IC 342
we find an age of $10^{6.8-7.8}$ years (B\"oker \etal
1999b). Unfortunately, most of the observed star clusters are much
fainter than that in IC~342, making this approach extremely
challenging. A better approach may therefore be to construct
population synthesis models for low resolution spectra, which are
easier to obtain at adequate S/N. We recently demonstrated the
feasibility of this approach for the nearby starburst galaxy NGC 4449
(B\"oker \etal 2001a). A ground-based optical echellette spectrum
shows strong Balmer absorption lines and a pronounced 4000 {\AA}
break, both reddening-free indicators of the strength of a young
population. Data-model comparison shows that the light of the cluster
is dominated by a very young (6-10 Myr) population of stars.  This
conclusion is supported by an independent analysis of broad-band
colors and a near-IR spectrum (Gelatt, Hunter \& Gallagher 2001).
Interestingly, many of the nuclear star clusters that have so far been
studied in detail turn out to be young. In addition to the two
galaxies already mentioned, our own Milky Way has a central cluster
that is only $\sim 10^{6.5}$ years old (Krabbe \etal 1995; Najarro
\etal 1997), and M31 and M33 have blue nuclei when viewed at HST
resolution that are quite possibly young star clusters (Lauer \etal
1998). However, it should be kept in mind that the existing studies
have probably been biased towards the brightest, and hence youngest,
clusters. A complete unbiased study is therefore called for. Such a
study will require the superior spatial resolution of HST, to cleanly
separate the cluster light (with a half-light radius of $\sim 0.1$
arcsec) from that of the underlying disk.  \looseness=-2

\subsection{Bars}

Bars are very common in spiral galaxies. Estimates from the RC3
suggest that 50--60\% of spiral galaxies are barred. More detailed
analysis of near-IR images suggests that the fraction of barred
galaxies may be as high as 60--80\% (e.g., Knapen, Shlosman \&
Peletier 2000). The Milky Way is now known to be barred as well (e.g.,
Kuijken 1995; Englmaier \& Gerhard 1999). Bars have the potential of
causing secular evolution in galaxy disks, and may be responsible for
the formation of bulges. Numerical simulations have shown that
dynamical resonances in the non-axisymmetric potential of a stellar
bar provide an efficient mechanism for the dissipation of angular
momentum and the subsequent infall of disk gas towards the galaxy
center (e.g. Athanassoula 1992a,b). Of particular interest in this
context is a model suggested by Friedli \& Benz (1993) and developed
further by Norman, Sellwood \& Hasan (1996). They pointed out that
build-up of a central mass concentration can dissolve stellar bars,
and lead to the formation of a bulge via collective bending
instabilities (Raha \etal 1991; Merritt \& Sellwood 1994). The Norman
\etal simulations show that only 5\% of the combined disk and bar mass
in a central concentration is sufficient to destroy a bar on very
short timescales, leading to a bulge-like distribution of stars. Both
a central star cluster and/or a central black hole (now believed to be
ubiquitous in galaxies; e.g., Kormendy \& Richstone 1995; Magorrian
\etal 1998; van der Marel 1999; Ferrarese \& Merritt 2000; Gebhardt
\etal 2000) could therefore potentially play a role in the destruction
of bars. Accurate statistics of nuclear star cluster and black hole
masses in spiral galaxies are required to address this
quantitatively. Note also that since stellar bars are an effective
funneling mechanism of gaseous matter into the central regions of
galaxies, the destruction of a stellar bar may eliminate further
matter infall towards the galaxy center. This may have implications
both for the feeding of an AGN, and for the continuing availability of
gas for the formation of new young star clusters.

\section{Elliptical Galaxies}

\subsection{Surface Brightness Cusps}

The high spatial resolution of HST has allowed astronomers to study
the photometric structure of elliptical galaxy nuclei with
unprecedented detail. The main result from a large number of different
studies (e.g., Lauer \etal 1995; Carollo \etal 1997) has been that at
the $\sim 0.1''$ resolution limit of HST, virtually all galaxies have
power-law surface brightness cusps, $I \propto r^{-\gamma}$, with
$\gamma>0$ and no observed transition to a homogeneous core. In
addition, the surface brightness profiles fall in two categories:
`core' profiles, which have a break at a resolved radius and a shallow
slope inside that radius, and `power-law' profiles, which have a steep
slope down to the resolution limit and no clear break (Faber \etal
1997). Core galaxies are defined to have $\gamma \leq 0.3$ at the HST
resolution limit, while power-law galaxies generally have $\gamma \geq
0.5$. The classification in two groups is motivated by the fact that
there appears to be a disjunct, rather than a continuous division in
profile properties, and by the fact that the overall physical
properties of core galaxies and power-law galaxies are
different. Among other things, the nuclear properties correlate with
luminosity. Galaxies with $M_V < -22$ have core profiles, galaxies
with $M_V > -20.5$ have power-law profiles, and both profile types
occur in galaxies with $-22 < M_V < -20.5$.

The physics underlying the observed cusp slopes and correlations
remains somewhat of a mystery. One possible interpretation is to
assume that the cusps are due to the presence of black holes. It has
long been know that the presence of a BH in the center of a stellar
system induces a power-law cusp in the mass distribution (e.g.,
Bahcall \& Wolf 1976; Young 1980). In fact, the adiabatic black hole
growth models of Young fit the observed cusp slopes quite well for
reasonable black hole masses (van der Marel 1999). However, this may
be somewhat fortuitous, given that the adiabatic black hole growth
models appear oversimplified. First, they assume that the black hole
grows slowly in a pre-existing stellar system. Models in which black
hole and galaxy formation occur simultaneously may be more realistic
(e.g., Haehnelt, Natarajan \& Rees 1998). Second, they assume that
galaxies initially form with homogeneous cores. There is evidence that
the formation of low-luminosity ellipticals involves considerable
dissipation (Bender, Burstein \& Faber 1992), and this may have caused
these galaxies to have a central cusp to begin with. And third,
galaxies are not expected to evolve in isolation, but may experience
mergers and accretion events. If both galaxies in a merger contain a
central black hole, then the evolution of the black hole binary will
lower the cusp slope near the center (Quinlan \& Hernquist 1997). It
has been suggested that this is the mechanism responsible for the
small cusp slopes in high-luminosity galaxies (Faber \etal
1997). However, it is not clear why this same mechanism wouldn't also
lead to small cusp slopes in the denser low-luminosity galaxies.  So
while considerable insight has been gained in the central stellar
density distribution of elliptical galaxies, much remains to be
understood about the underlying physics.

\subsection{Nuclear Disks}

With HST it has been possible to study the dust and ionized gas in
early-type galaxies at $\sim 10$ times better resolution than
before. Interestingly, this has revealed the existence of previously
unknown structures: `nuclear disks', with typical sizes of 0.1--1
kpc. In some cases the disks are seen only in ionized gas emission
(e.g., M87; Ford \etal 1994), while in other cases they have strong
associated dust absorption, visible as a dust lane across the nucleus
(e.g., NGC 7052; van der Marel \& van den Bosch 1998; see also
Fig.~1b.). The observed morphology and kinematics of the nuclear disks
imply approximately circular motion under the influence of gravity,
and they have therefore been used successfully to measure the masses
of central black holes. The nuclear disks are sometimes kinematically
decoupled from the stellar body. In NGC 4261 the stars rotate around
the major axis (indicating a prolate geometry), while the nuclear dust
and gas disk rotates around the minor axis (Ferrarese, Ford \& Jaffe
1996). This leads to a picture in which gas and dust have an external
origin, due to some accretion or merger event, and have settled at
small radii in one of the symmetry planes of the galactic
potential. HST observations show nuclear dust in $\sim 80$\% of
early-type galaxies (van Dokkum \& Franx 1995). The dust morphology is
often patchy, and not always in the form of a disk. The detection rate
of nuclear dust in radio-loud systems is roughly twice that in
radio-quiet systems. This indicates a possible relation between the
dust and BH fueling.

Small stellar disks are also sometimes found, examples being the
nearly edge-on disks in NGC 4342, NGC 4570 (van den Bosch, Jaffe \&
van der Marel 1998) and NGC 3115 (Kormendy \etal 1996).  A plausible
formation scenario for these disks is the dissipative infall of gas,
which settles into a disk and forms stars (Hernquist \& Barnes
1991). In such a scenario kinematical decoupling (e.g., Mehlert \etal
1997) arises naturally, because the gas maintains its own angular
momentum. On the other hand, strong color and stellar population
differences between the kinematically decoupled core and the main body
are the exception rather than the rule (Carollo \etal 1997),
indicating that the infall may have occurred at early times.

\section{Galaxy Mergers and Merger Remnants}

\subsection{Starbursts and ULIRGs}

Numerical simulations show that torques and shocks during a merger
efficiently remove angular momentum from the gas in the merging
galaxies, which therefore flows to the galaxy center(s) (e.g., Mihos
\& Hernquist 1996). The gas gets compressed, and is thus expected to
form stars. This is generally confirmed observationally; e.g., HST
observations of the well-known Antennae galaxies beautifully reveal
wide-spread star and star cluster formation (Whitmore \etal
1999). Often the star formation is dust-enshrouded, in which case the
majority of the energy is emitted in the IR. In the most spectacular
cases this leads to Ultra-Luminous Infra-Red Galaxies (ULIRGs). Such
galaxies are almost always seen to be interacting systems or merger
remnants, and HST observations have shown that in $\sim 20$\% of
ULIRGs more than two galaxies appear to be involved in the interaction
(Borne \etal 2000). The gas flow towards the center in an interacting
system may also fuel an AGN, which can also contribute to the observed
IR emission. In some cases the presence of an (obscured) AGN is
unambiguous from the observation of strong X-ray emission, as in e.g.,
NGC 6240 (e.g., Vignati \etal 1999; an HST image is shown in
Fig.~1c). Genzel \etal (1998) have used IR emission-line diagnostic
ratios from ISO data to address the nature of the energy source of
ULIRGs more quantitatively. They find that 70--80\% of the ULIRGs in
their sample are predominantly powered by recently formed massive
stars, and 20--30\% by a central AGN. At least half of the sources
probably have both an active nucleus and starburst activity in a 1--2
kpc diameter circumnuclear disk/ring.

\subsection{Formation of ellipticals from mergers}

The possibility that many elliptical galaxies formed from mergers of
disk galaxies is a topic of continuing interest. That mergers form
elliptical-like remnants has been demonstrated through numerical
simulations (e.g., Hernquist 1992), and ground-based imaging has shown
that many merger remnants have r$^{1/4}$ luminosity profiles (e.g.,
Stanford \& Bushouse 1991). These arguments, along with the detection
of shells, ripples, dust disks and kinematically decoupled cores in
elliptical galaxies, support this `merger hypothesis' (e.g.,
Kennicutt, Schweizer \& Barnes 1998).

Theoretical arguments indicate that it is in the nuclei of remnants
where the merger hypothesis may face its most stringent test. If
dynamical relaxation is the dominant physical process in mergers, then
remnant nuclei will be very diffuse with large cores (Hernquist 1992),
unless the progenitor nuclei were dense to begin with. If both merging
galaxies contain a central black hole, then the stellar density of the
merger remnant will be lower than that of the progenitor galaxies
(Quinlan \& Hernquist 1997).  Alternatively, if mergers are
accompanied by strong gaseous dissipation and central starbursts, then
the remnant may have a high stellar density and steep luminosity
profile (Mihos \& Hernquist 1994).

A comparison between the observed nuclear properties of merger
remnants and elliptical galaxies can shed more light on the viability
of the merger hypothesis and on the physical processes that govern the
structure of merger remnants. We have undertaken a study with
HST/NICMOS to address this issue. Preliminary results were presented
in van der Marel \& Zurek (2000). Fourteen late-stage merger remnants
(identified morphologically) were imaged in $J$, $H$ and $K$, and for
each galaxy the surface brightness profile was inferred from the
data. The profiles generally show a central cusp, as do elliptical
galaxies. Most of the galaxies in the sample have cusp slopes that are
similar to those typical for ellipticals of the same luminosity. On
the other hand, three of the fourteen galaxies stand out by having a
higher central luminosity density than typical for elliptical
galaxies. In two of these galaxies the light becomes bluer towards the
center, presumably due to recent star formation. It is therefore
plausible to attribute the excess luminosity density to stars formed
recently from gas that has fallen towards the center, as predicted in
the models of Mihos \& Hernquist (1994). The young stars will fade
with time, so that these galaxies will be more similar to elliptical
galaxies after another few Gyrs. So these results are not inconsistent
with the hypothesis that merging disk galaxies form
ellipticals. However, what fraction of ellipticals in the Universe
formed from a major merger remains an open question.

\bigskip

It is a pleasure to thank Torsten B\"oker for assistance with the text
of Section~2. Several of the HST projects discussed in this review are
supported by grants awarded by the Space Telescope Science Institute
which is operated by the Association of Universities for Research in
Astronomy, Incorporated, under NASA contract NAS5-26555.

\end{document}